\documentclass[aps,pre,twocolumn,showpacs,superscriptaddress,preprintnumbers,amsmath,amssymb]{revtex4}
\usepackage{amsmath}
\usepackage{graphicx}
\usepackage{dcolumn}
\usepackage{bm}
\usepackage{float}

\begin{document}

\title{Modification of the formation of high-Mach number electrostatic 
shock-like structures by the ion acoustic instability}


\author{M. E. Dieckmann}\thanks{Electronic mail: Mark.E.Dieckmann@itn.liu.se}
\affiliation{Dept of Science and Technology, Link\"oping University,
SE-60174 Norrk\"oping, Sweden}

\author{G. Sarri}
\affiliation{Centre for Plasma Physics, School of Mathematics and Physics,  
Queen's University of Belfast, Belfast BT7 1NN, United Kingdom}

\author{D. Doria}
\affiliation{Centre for Plasma Physics, School of Mathematics and Physics,  
Queen's University of Belfast, Belfast BT7 1NN, United Kingdom}

\author{M. Pohl}
\affiliation{Univ Potsdam, Inst Phys \& Astron, D-14476 Potsdam, Germany}
\affiliation{DESY, D-15738 Zeuthen, Germany}

\author{M. Borghesi}
\affiliation{Centre for Plasma Physics, School of Mathematics and Physics,  
Queen's University of Belfast, Belfast BT7 1NN, United Kingdom}


\date{\today}


\begin{abstract}
The formation of unmagnetized electrostatic shock-like structures with a high Mach number is examined with one- and two-dimensional particle-in-cell (PIC) simulations. The structures are generated through the collision of two identical plasma clouds, which consist of equally hot electrons and ions with a mass ratio of 250. The Mach number of the collision speed with respect to the initial ion acoustic speed of the plasma is set to 4.6. This high Mach number delays the formation of such structures by tens of inverse ion plasma frequencies. A pair of stable shock-like structures is observed after this time in the 1D simulation, which gradually evolve into electrostatic shocks. The ion acoustic instability, which can develop in the 2D simulation but not in the 1D one, competes with the nonlinear process that gives rise to these structures. The oblique ion acoustic waves fragment their electric field. The transition layer, across which the bulk of the ions change their speed, widens and their speed change is reduced. Double layer-shock hybrid structures develop.  
\end{abstract}
\maketitle

\section{Introduction}

Collision-less plasma shocks are ubiquitous in the dilute solar system
plasmas and in astrophysical plasmas. Their internal structure is 
fundamentally different from their collisional counterparts, which behave
similarly to shocks in gases. Collisional shocks can transform almost
instantly the directed flow energy of the incoming upstream plasma into 
heat by means of binary collisions between the plasma particles. Particle
beams are rapidly thermalized and the plasma can be described by a unique
temperature value at any position. In the case of collision-less plasma 
shocks, the upstream plasma is slowed down and heated up by electromagnetic 
fields as it crosses the shock boundary. Multiple plasma beams can be present 
at any location and it is possible that a subset of particles is accelerated 
to high energies by the shock while the bulk of the particles is thermalized. 
The structure of collision-less shocks depends strongly on the local plasma 
parameters, in particular on the background magnetic field, on the electron 
and ion temperatures and on the ion composition. A background magnetic field 
is particularly important, because it determines the wave mode that mediates 
the shock. 

The key role held by the background magnetic field is evidenced by the Earth's 
bow shock, which develops where the solar wind encounters the Earth's magnetic 
field. The relative speed between the solar wind and the Earth's magnetic 
field exceeds the ion acoustic speed and the Alfv\'en speed; the boundary 
separating the solar wind plasma and the magnetosheath's plasma is thus a 
shock \cite{Sckopke}. In spite of its low amplitude of about 5nT 
\cite{SolarWind}, the magnetic field of the solar wind assumes a vital role 
in determining the structure of the bow shock. If the solar wind's magnetic 
field is oriented perpendicularly \cite{QuasiPerp} to the shock's normal, 
the shock transition layer is narrow. As the angle between the magnetic 
field and the shock normal decreases, the shock transition layer widens
\cite{QuasiPar}. The shock boundary changes into a train of SLAMS (short 
large amplitude magnetic structures) for small angles \cite{SLAMS}.

The most basic type of shock develops in unmagnetized plasma. Such shocks 
have been observed in a wide range of experiments, e.g. 
\cite{Exp1,Exp2,Exp3,Exp4,Exp5,Exp6}, they have been addressed theoretically 
\cite{Bardotti,Sorasio,Raadu,Yannis,Antoine} and by means of numerical 
particle-in-cell (PIC) and hybrid simulations 
\cite{ForslundA,ForslundB,Karimabadi,Kato,Parametric}.
The shock is sustained by the electrostatic field that is tied to the density
gradient between the downstream and upstream plasmas. This density gradient
results in turn from the slow-down of the upstream ions by the electrostatic
field as they cross the shock transition layer. The electric field and the 
plasma compression are thus conjoined processes. The ambipolar electrostatic 
field is a consequence of the different electron and ion mobilities. 
Electrons can escape from the denser downstream plasma into the upstream 
plasma. A positive net charge develops in the downstream plasma and a 
negative one in the upstream plasma. The space charge results in an 
electrostatic field across the shock that helps confining the downstream 
electrons. A shock forms if this electric field is strong enough to slow 
down the incoming upstream ions to a speed in the downstream reference frame, 
which is comparable to the downstream ion's thermal speed. This condition 
imposes an upper limit on the speed, or more specifically on the Mach number, 
of non-relativistic and unmagnetized collision-less shocks.

Here we examine by means of PIC simulations the formation of electrostatic
structures out of the collision of two equal and spatially uniform plasma 
clouds at a contact boundary, which is orthogonal to the collision direction. 
Each cloud consists of one electron and one ion species. The electrons and 
ions of each cloud have the same density, the same temperature and the same 
mean speed at the simulation's start. The plasma is thus free of net charge 
and current and initially all electromagnetic field components are set to 
zero. No particles are introduced after the simulation has started. The Mach 
number, which corresponds to the collision speed between both clouds, is close 
to the maximum one, which resulted in the formation of electrostatic shock-like
structures in similar simulations \cite{Parametric}. These shock-like
structures can at least initially not be classified \cite{GreatPaper} as 
electrostatic shocks due to transient effects, which arise from our choice of 
initial conditions. The shock-like structures tend to form slowly for high 
Mach numbers of the collision speed, which allows for the simultaneous 
development of the ion acoustic instability between counter-streaming ion 
beams \cite{Karimabadi,Kato,ForslundC}. It has been shown recently that the 
ion acoustic instability can destabilize an already existing electrostatic 
shock \cite{Kato}. Here we examine this instability as it develops already 
during the formation phase of a shock. Our results are as follows. 

Our first simulation study resolves only the direction that is aligned with 
the relative velocity vector between both clouds. This geometry excludes the 
ion acoustic instability for the considered initial conditions. The simulation 
confirms that the formation time of the shock-like structures is delayed by 
the large collision speed; the electrostatic fields that mediate these 
structures grow slowly. They need several tens of inverse ion plasma 
frequencies to reach the amplitude, which is necessary to let the 
counter-streaming ion beams collapse into a pair of shock-like structures.
This delay is comparable to the one observed in Ref. \cite{Parametric} for 
a similar collision Mach number and for ions with a charge-to-mass ratio
that is 2/3 of the one used here, suggesting that the peak Mach number of 
such structures may not depend strongly on the value chosen for this ratio. 
The latter can have a significant impact on the shock formation for faster 
collisions \cite{MassRatio}. These shock-like structures gradually evolve
into electrostatic shocks as they separate. The forward and reverse shocks 
are time-stationary in their rest frame in the 1D simulation and they 
propagate at a constant speed, as in previous one-dimensional PIC simulation 
studies \cite{Parametric}. 

Our 2D simulation study employs initial conditions that are identical
to those of the first one and it has the purpose to assess the impact of 
the ion acoustic instability, which is observed in the context of laser
plasma experiments \cite{Bulanov}, on the shock formation. This instability 
develops between two counterstreaming ion beams if their relative speed is 
significantly less than the thermal speed of the electrons. The ion acoustic 
waves can only grow if the projection of the beam velocity vector onto the 
direction of the wave vector yields a sub-sonic speed modulus. This 
constraint implies for our initial conditions that the waves must move 
obliquely to the beam velocity vector \cite{ForslundC}, which requires a 
2D simulation geometry. We observe that the electric field of the shock-like
structures and the one due to the ion acoustic instability develop 
simultaneously and eventually reach a comparable amplitude. The ion acoustic 
waves fragment the shock's electric field altering the balance between the 
downstream pressure, which has contributions by ram pressure and thermal 
pressure, and the pressure of the incoming upstream plasma that sustains 
the shock-like structure. The velocity change of the bulk of the inflowing 
ions is comparable to the ion acoustic speed and, thus, well below that 
observed in the 1D simulation. We observe a widening of the transition layer, 
across which the ions change their speed as they move from the upstream to 
the downstream region.

A comparison of the electron velocity distributions downstream of the shocks
computed by the 1D and 2D simulations suggests that the flat-top distribution,
which is observed in the 1D simulation and in Ref. \cite{Parametric}, 
results from the reduced simulation geometry. A pronounced maximum of the
velocity distribution function develops at low speeds in the 2D simulation 
and the distribution function gradually decreases with increasing speed 
moduli. We attribute the modified velocity distribution function to the 
interaction of electrons with the strong ion acoustic waves.

The structure of our manuscript is as follows. Section 2 describes
qualitatively how an electrostatic shock forms, it summarizes the numerical 
scheme of a PIC code and it details our initial plasma conditions. Section 3 
presents the simulation results and section 4 is the discussion.
 
\section{Initial conditions and the simulation method}

\subsection{The shock model}

\begin{figure}
\includegraphics[width=8.2cm]{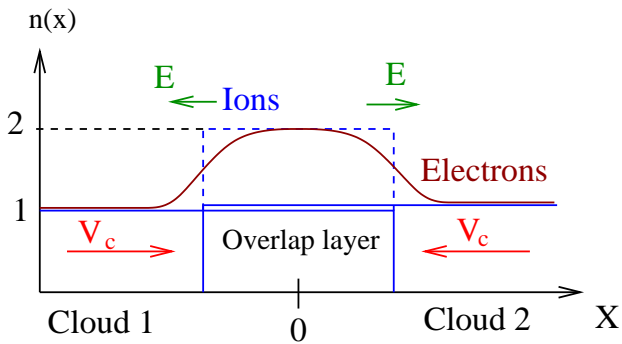}
\caption{Shock formation: Two equal plasma clouds consisting of electrons
and ions, each with the density $n_0=1$, collided initially at the position 
$x=0$ at the speed $2v_c$. The figure shows the system a short time after 
the collision, when clouds 1 and 2 have interpenetrated for a short distance. 
The ion density in this overlap layer is $n(x)=2$. Some electrons stream out 
of this layer due to their high mobility and the resulting net charge puts 
the overlap layer on a positive potential relative to the surrounding plasma 
clouds.}\label{fig1}
\end{figure}

Non-relativistic electrostatic and unmagnetized shocks form due to 
the ambipolar electric field of a plasma density gradient and are
stabilized by it. Figure \ref{fig1} illustrates this mechanism assuming
that the ions are cool.
Two plasma clouds, each consisting of electrons and ions, collide initially 
at the position $x=0$. The ions and electrons of each cloud move at the 
equal mean speed modulus $v_c$ towards $x=0$. The density of the electrons 
and of the singly charged ions is $n_0$ and each plasma cloud is thus 
initially free of any net charge and current. The low thermal speed of the 
ions preserves their number density distribution on electron time scales. 
The ion number density in the overlap layer is thus initially $2n_0$ 
and it decreases to $n_0$ at the two boundaries between the overlap layer 
and both incoming plasma clouds. Some electrons diffuse across the boundaries, 
leaving behind a positively charged overlap layer. The overlap layer goes on 
a positive potential relative to both clouds, which is independent of $v_c$. 
The associated unipolar electric field at each of the boundaries points 
towards the incoming plasma clouds. It thus confines electrons to the overlap 
layer, it results in an expansion of ions from the overlap layer and in a 
slow-down of the ions of the incoming plasma clouds as they cross the overlap 
layer's boundary. 
 
The evolution of the overlap layer is determined by how the kinetic energy of 
the incoming ions in the reference frame of the overlap layer compares to the 
potential energy they gain as they enter the overlap layer. If the kinetic 
energy is significantly larger, the ions of both clouds overcome the positive 
potential of the overlap layer and the counterstreaming ions thermalize via 
beam instabilities. Otherwise, the evolution of the overlap layer depends on 
how the pressure of the plasma in the overlap layer compares to the pressure 
that is excerted on its boundary by the incoming plasma. This balance is 
mediated by the ambipolar electric field. The overlap layer expands in the 
form of a rarefaction wave \cite{Sack}, if its pressure can not be balanced 
by the pressure of the upstream plasma. A shock solution can exist if the 
pressure of the overlap layer and of the upstream plasma are equal in some
reference frame. The shock is stationary in this frame, which is henceforth 
denoted as the shock frame. The ram pressure dominates the upstream plasma 
pressure in this frame and the thermal pressure contributes most to that of 
the downstream plasma.

The formation of an electrostatic shock is an inherently non-linear process 
that does not depend on wave and beam instabilities for the low Mach number
of the collision speed, which we consider here. This is demonstrated by our 
1D simulation, where the ion beam instability is excluded by the simulation 
geometry while the Buneman instability \cite{Buneman,Amano} is suppressed by 
the large thermal speed of the electrons. The slow-down of the incoming ions 
in the reference frame of the overlap layer is tied to a density increase 
via the continuity equation. The ion density in the overlap layer increases 
beyond $2n_0$ and the potential difference between the compressed overlap 
layer and the incoming plasma cloud increases accordingly. The larger 
potential difference results in an even stronger slow-down and compression 
of the incoming ions. This non-linear and self-amplifying process, which has 
been resolved experimentally \cite{Exp6}, is eventually halted by the 
formation of a shock. The shock separates the downstream region, which is 
the compressed overlap layer, from the upstream region. The latter corresponds 
to the incoming unperturbed plasma cloud. The frequently observed partial 
reflection of the incoming ions by the shock potential 
\cite{ForslundA,ForslundB} gives rise to a foreshock region that is occupied 
by the incoming plasma cloud and by a beam of shock-reflected ions. 

\subsection{The particle-in-cell method and the initial conditions}

The particle-in-cell (PIC) method approximates the plasma by an ensemble of
computational particles (CPs) and the collective electromagnetic fields 
$\mathbf{E}$ and $\mathbf{B}$ are computed on a numerical grid. These 
fields are generated by the current- and charge density distributions 
$\mathbf{j}(\mathbf{x},t)$ and $\rho (\mathbf{x},t)$ in the plasma. 
The electromagnetic fields are evolved in time by Amp\`ere's and Faraday's 
laws, 
\begin{eqnarray}
\nabla \times \mathbf{B} = \mu_0 \mathbf{j} + \mu_0 \epsilon_0 \partial_t 
\mathbf{E}, \\
\nabla \times \mathbf{E} = -\partial_t \mathbf{B},
\end{eqnarray}
which are discretized and represented on a numerical grid. Gauss' law is 
either fulfilled as a constraint or through a correction step while $\nabla 
\cdot \mathbf{B} = 0$ is usually preserved to round-off precision.

Each CP is characterized by a charge $q_j$ and mass $m_j$, by a position 
vector $\mathbf{x}_i$ and by a velocity vector $\mathbf{v}_i$. The subscript 
denotes the $i^{th}$ CP of the ensemble that represents the plasma species 
$j$. The ratio $q_j / m_j$ must be equal to that of the approximated plasma
species, which can be electrons, positrons or ions. The relativistic
momentum $\mathbf{p}_i$ of each CPs is evolved in time with a discretized 
form of the Lorentz force equation $d\mathbf{p}_i / dt = q_j \left ( 
\mathbf{E}(\mathbf{x}_i)+ \mathbf{v}_i \times \mathbf{B}(\mathbf{x}_i) 
\right )$. The momentum of the CP is $\mathbf{p}_i = m_j \Gamma_i 
\mathbf{v}_i $ and $\Gamma_i$ is its relativistic factor. The position is 
updated with $\mathbf{v}_i$ and the simulation time step. The electromagnetic 
fields in the Lorentz force equation have been interpolated from the grid to 
the position of the CP. The charge and current contributions of each CP are 
interpolated back to the grid. The contributions of all CPs are summed up 
to give $\rho (\mathbf{x})$ and $\mathbf{j}(\mathbf{x})$, 
which are used to update the electromagnetic fields on the grid.

The ensemble properties of the CPs are close to those of a true plasma 
provided that the numerical resolution is adequate. The CPs interact via 
the collective electromagnetic fields, while binary collisions are usually 
neglected. PIC codes can represent all kinetic wave modes and processes 
captured by the Vlasov-Maxwell set of equations \cite{Dupree}, provided 
that the numerical resolution is appropriate. An in-depth description of 
the PIC method can be found elsewhere \cite{Dawson}. We use here the TwoDem 
code that is based on the virtual particle-mesh method \cite{OldPIC}. The 
code solves the relativistic equations of motion for the CPs. Our initial 
conditions imply however that all velocities stay non-relativistic. 

We perform two simulations, which use the same initial conditions for the
plasma. The simulation box with length $L$ is subdivided along the 
x-direction. Plasma cloud 1 is placed in the interval $-L/2 \le x < 0$
and the interval $0 < x \le L/2$ is occupied by the plasma cloud 2. Each 
cloud is composed of one electron species and one species of singly charged 
ions. Both have the number density $n_0$, which defines the electron plasma
frequency $\omega_{pe}={(n_0 e^2 / m_e \epsilon_0)}^{1/2}$. The 
ion-to-electron mass ratio is set to $m_i / m_e =$ 250, giving an ion plasma
frequency $\omega_{pi}=\omega_{pe}/250^{1/2}$. The spatially uniform 
electrons and ions have a Maxwellian velocity distribution with the 
temperature 10 eV. The electron thermal speed is $v_e = 1.325 \times 10^6$ 
m/s and that of the ions is $v_i = v_e / 250^{1/2}$. The electrons and ions 
of each cloud move at the speed $v_c = 3 \times 10^5$ m/s towards $x=0$. 
The low collision speed $2v_c / v_e \approx 0.45$ suppresses the Buneman 
instability between the ions of one cloud and the electrons of the second 
cloud. 

We define the ion acoustic speed $v_s$ through $v_s^2 = \gamma_s k_B (T_e 
+ T_i) / m_i$. This speed is meaningful in a fluid model, where collisions 
enforce a single Maxwellian velocity distribution and, thus, a single 
temperature for electrons and for each ion species at any given position
and where Landau damping is absent. The ion acoustic waves are Landau damped 
in a kinetic collision-less framework unless the electrons are much hotter 
than the ions. Multiple beams of particles of a single species can be present 
at the same location and the velocity distribution is not necessarily a 
Maxwellian one. The ion acoustic speed and the shock's Mach number are thus 
not as meaningful in a collision-less plasma as they are in a fluid model. 
We introduce the ion acoustic speed here to compare our initial conditions, 
which involve Maxwellian velocity distributions for one electron and one ion 
species at each point in space, to those in related simulation studies and 
to the conditions found in laser-generated or astrophysical plasma. We assume 
that both species have the same adiabatic constant $\gamma_s = 5/3$, which 
gives us the Mach number of the collision speed $v_c / v_s \approx 2.3$.

The 1D simulation resolves the x-direction by 3000 simulation grid cells of
size $\Delta_x = 0.95 \lambda_D$, where the Debye length $\lambda_D = v_e / 
\omega_{pe}$. Electrons and ions are each represented by 4464 CPs per 
cell. The 1D simulation resolves a time interval $t\omega_{pi}=157$. The 2D 
simulation employs 2500 grid cells along the x-direction and 300 grid cells 
along the y-direction. The cell size $\Delta_x = \Delta_y = 0.95 \lambda_D$. 
Electrons and ions are each represented by 160 CPs per cell. We employ 
periodic boundary conditions and we do not introduce new particles after the 
simulations have started. The two colliding electron-ion clouds are thus the 
only plasma constituents throughout the simulation. The back ends of the 
plasma clouds detach from the boundaries in the x-direction and move towards 
the center of the box. The 2D simulation covers a time interval $t\omega_{pi}
=86$ and in this simulation $t v_c \approx L/8$. The simulations are thus 
stopped long before the front of one plasma cloud reaches the back end of the 
counter-streaming second plasma cloud. 

\section{The simulation results}

In what follows we present the results of our 1D and 2D simulations. The 
electric field amplitude is expressed in units of $\omega_{pe} m_e c / e $,
space in units of the electron Debye length $\lambda_D$ and time in units 
of $\omega_{pi}^{-1}$. 

\subsection{The 1D simulation}

\begin{figure}
\includegraphics[width=8.2cm]{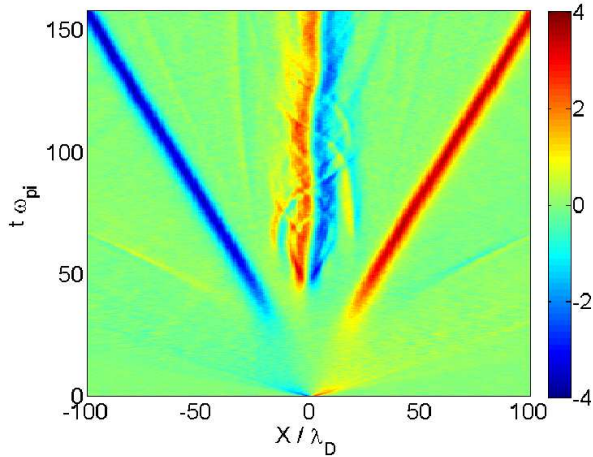}
\caption{The spatio-temporal electric field distribution in the 1D simulation:
The color corresponds to $10^3 E_x$, space is given in units of the
electron Debye length $\lambda_D$ and time is normalized to the ion plasma
frequency $\omega_{pi}$.}
\label{fig2}
\end{figure}

Figure \ref{fig2} shows the spatio-temporal evolution of the electric field 
in the 1D simulation, which can be subdivided into three intervals. The first 
interval $t\omega_{pi} < 5$ corresponds to a shock-less interpenetration of 
both plasma clouds, as depicted in Fig. \ref{fig1}. Strong electric fields 
are observed in the spatial interval $-5 < x/\lambda_D < 5$ during this time. 
The ion density gradient at both boundaries of the overlap layer is large, 
resulting in a strong ambipolar electrostatic field. The ion density gradient 
is eroded in time due to ion diffusion, which is a consequence of the ion's 
thermal velocity spread. The electric field amplitude decreases accordingly 
and it spreads out in space. The potential difference between the overlap 
layer and the incoming plasma clouds remains unchanged though, because it is 
determined by the difference in the positive charge density $\approx en_0$ 
between the overlap layer and the incoming plasma cloud and by the electron 
temperature.

The second time interval between $5 < t \omega_{pi} < 30$ is characterized
by a broad distribution of weak electric fields that seem to maintain a 
constant amplitude. The positive potential of the overlap layer is not 
capable of slowing down the ions of both incoming plasma clouds to a speed 
in the rest frame of the overlap layer that is comparable to the ion thermal 
speed; no shock develops. A lower value of $v_c$ would result in their 
formation on electron time scales. However, the potential of the overlap 
layer in the 1D simulation slows down and compresses the incoming ions close 
to the boundary and the ion density is increased locally beyond $2n_0$. The 
positive potential within the overlap layer and, thus, the ion compression 
increase. The ion accumulation takes place at the boundary between the overlap 
layer and the incoming plasma cloud if the ions are cold. The thermal 
diffusion of warm ions implies though that this boundary spreads out. The ion 
compression beyond the density $2n_0$ is achieved in this case at the 
location, which corresponds to the maximum of the electrostatic potential. 

The coupling between the ion slow-down and the increase of the electrostatic 
potential implies that this is a self-amplifying process. In what follows we 
refer to this instability as the ion compression instability. Eventually the 
potential difference between the compressed overlap layer and the incoming 
plasma is large enough to let both ion density accumulations collapse into 
shock-like structures during the time $40 < t \omega_{pi} < 50$. 

We observe two electric field pulses in the third time interval $t\omega_{pi}
>50$, which are propagating away from $x=0$ at a constant speed. Their 
propagation speed in the reference frame of the simulation box can be 
estimated from Fig. \ref{fig2} to be $|v_p| \approx 80 \lambda_D /(110 
\omega_{pi}^{-1})$ or $|v_p| / v_s \approx 0.3$. Their Mach number in the 
reference frame of the incoming plasma cloud and camputed with respect to 
the initial ion acoustic speed is $M_s \approx 2.6$, since $v_c / v_s 
\approx 2.3$. This Mach number
and the formation time are similar to the ones of the fastest collision in 
Ref. \cite{Parametric}, which resulted in shocks. The electric field 
demarcates the transition layer of the shock-like structure, which has here 
a width of about 10 $\lambda_D$. A bipolar electric field structure is 
present at $x\approx 0$. The polarization of this field distribution implies 
that a negative excess charge is present at $x\approx 0$, which is typical 
for an ion phase space hole \cite{IonHole}.

\begin{figure}
\includegraphics[width=8.2cm]{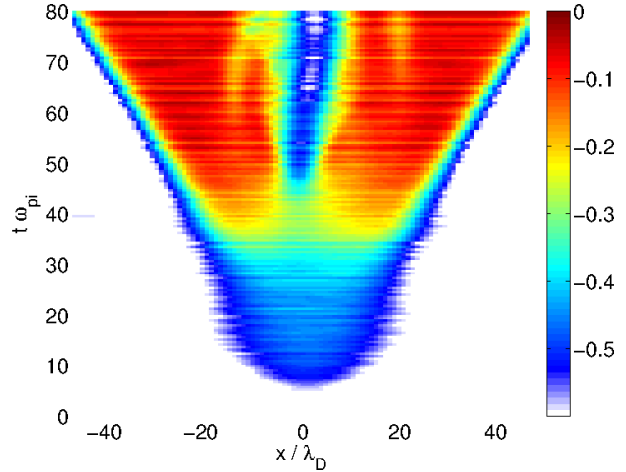}
\caption{The normalized electrostatic potential $\tilde{U}(x)$.}\label{fig3}
\end{figure}

We compute the potential $U(k\Delta_x)$ at the cell $k$ from the electric 
field distribution (Fig. \ref{fig2}) through the integration $U(k\Delta_x) 
=- \sum_{i=1}^{k} E_x (i\Delta_x ) \, \Delta_x$, where all quantities are
given here in their unnormalized SI units. The cell with the index $i=1$ 
corresponds to the left boundary. We express the potential $U$ in units of 
$E_k / e$ with $E_k = m_i {(2.6 v_c / 2.3)}^2 / 2$. This is the kinetic 
energy of an ion in the reference frame of the electric pulse, which moves 
towards the pulse at the speed $v_c$ in the box frame. The mean value of the 
fully developed potential is subtracted. The potential $\tilde{U}$ in this 
normalization is shown in 
Fig. \ref{fig3}. 
It grows first at $x\approx 0$ and reaches a pratically stationary 
distribution between $10< t\omega_{pi} < 30$. It grows to larger values at 
$t\omega_{pi} \approx 40$ and at $|x|/\lambda_D \approx 20$. This is well 
behind the positions $|x|/\lambda_D = 40 v_c / (\omega_{pi} \lambda_D) 
\approx 150$ that would be reached by ions with the speed modulus $v_c$ that 
moved away from the position $x=0$ at $t=0$. The potential depletion at 
$x\approx 0$ forms together with the pair of electric field pulses.

\begin{figure}
\includegraphics[width=8.2cm]{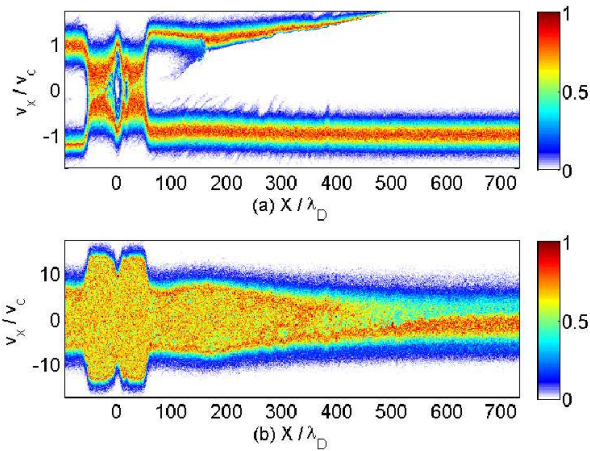}
\caption{The phase space distributions $f_{i,e}(x,v_x)$ from the 1D simulation
at the time $t\omega_{pi}=86$: Panel (a) shows the ion distribution and panel 
(b) shows the electron distribution. Space and velocity are expressed in 
units of the Debye length $\lambda_D$ and of the initial cloud speed $v_c$. 
The density is normalized to its peak value and displayed on a linear color 
scale (enhanced online).}\label{fig4}
\end{figure}

Figure \ref{fig4} shows the plasma phase space distribution at the time
$t\omega_{pi}=86$ when the pair of electric field pulses and the potential 
depletion at $x\approx 0$ have fully developed (See Fig. \ref{fig3}). The 
online enhancement of Fig. \ref{fig4} animates the time evolution of the 
phase space density for $0\le t\omega_{pi}\le 157$. It visualizes the ion 
compression instability at the simulation's start, which is characterized 
by a gradual slow-down of the ions in the overlap layer. We focus in Fig. 
\ref{fig4} and in its online enhancement on the interval around the (forward) 
shock-like structure that moves towards increasing values of $x$.  
Figure \ref{fig4}(a) reveals the presence of shock-like structures at the 
positions $|x|/\lambda_D \approx 50$, which coincide with those of the 
strong unipolar electrostatic fields in Fig. \ref{fig2}. A single ion 
population with a non-Maxwellian velocity distribution is observed in most 
of the downstream region between both shock-like structures. The only 
exception is the ion phase space hole, which is located at $x\approx 0$ and 
gives rise to the bipolar electric field in Fig. \ref{fig2}.

The ion beam at $x/\lambda_D > 50$ and $v_x > 0$ shows two distinct phase 
space distributions. The phase space distribution in the interval $150 < 
x/\lambda_d < 500$ is that of the ion beam that crossed the overlap layer 
before the shock-like structures formed. The phase space profile of this 
beam section is that of a rarefaction wave \cite{SarriPRL}, which moves 
relative to the simulation frame of reference. The ions in the phase space 
interval $50 < x/\lambda_d < 150$ and $v_c>0$ consist of two ion populations, 
which can be seen most easily from the online enhancement of Fig. \ref{fig4}. 
The source of the faster ions is the downstream plasma. These ions have been 
accelerated in the upstream direction by the electric pulse. The slower 
ions with $v_x \approx v_c$ originate from the incoming plasma cloud. They 
have been reflected by the shock-like structure. An incoming ion with $v_x = 
-0.5v_c$ at $x/\lambda_D \approx 60$, which is reflected specularly by a 
shock that moves in the simulation frame at the speed $v_p \approx 0.3 v_c$ 
(See Fig. \ref{fig2}), moves back upstream at the speed $v_x / v_c \approx 
1.1$. 

The fact that the ions of this beam arise from the upstream population and 
the downstream population implies that the structure at $x/\lambda_D\approx 
50$ is not a pure electrostatic shock in the definition of Ref. 
\cite{GreatPaper}. An electrostatic shock is composed at best of two distinct 
ion populations; one population of trapped ions and one population of free 
ions, which move both from the low potential side ($x/\lambda_D > 50$ in Fig. 
\ref{fig4}(a)) to the high potential side. The free 
ions of the shock-like structure at $x/\lambda_D \approx 50$ correspond to 
the beam of incoming ions with $v_x < 0$. The ions are slowed down as they 
cross the structure. The incoming ions, which have been reflected by the 
shock-like structure, form the trapped population. However, we also find a 
second population of free ions: those that cross the structure at 
$x/\lambda_D = 50$ and move to increasing values of $x$. Ions that flow from 
the high-potential side to the low-potential side indicate a double layer. 
According to the classification in Ref. \cite{GreatPaper}, the structure at 
$x/\lambda_D \approx 50$ and, by symmetry, the one at $x/\lambda_D \approx 
-50$ are hybrid structures. Hence we refer to them as shock-like structures.

The double layer component of the shock-like structure at $x/\lambda_D 
\approx 50$ is strong in Fig. \ref{fig4}(a) because a dense population of 
ions, which correspond to the free ions that move from the left ($v_x > 0$) 
towards the shock-like structure at $x/\lambda_D \approx -50$ and traverse
the downstream region, reaches the right-moving structure. This is a transient 
effect. Once the downstream region between both structures is sufficiently 
wide to thermalize the downstream ions, the ion velocity distribution enclosed
by both shock-like structures will change into a Maxwellian one centered at 
$v_x=0$. The number density of the ions, which are fast enough to reach both 
shock-like structures and feed the double layer, will be much lower. The 
hybrid structure will change into an electrostatic shock. 

Figure \ref{fig4}(b) displays the electron distribution at $t\omega_{pi}=86$.
We can subdivide this distribution into three spatial intervals. The electron
distribution close to $|x|/\lambda_D \approx 700$ corresponds to the initial
distribution. The velocity distribution is close to a Maxwellian with a 
maximum that is shifted by $-v_c$. A large circular structure is observed in 
the displayed interval $x/\lambda_D < 400$. The increased positive potential, 
which results from the ion accumulation in this interval, confines the 
electrons. The trapped electrons move on closed phase space orbits. This
trapped electron population is a pre-requisite for double layers and shocks
\cite{GreatPaper}. The velocity distribution within this phase space structure 
is not Maxwellian but has a phase space density that is constant apart from 
statistical noise.

The small circular phase space intervals with a reduced electron density in 
this large cloud of trapped electrons are electron phase space holes. They 
are stable electrostatic structures in a 1D geometry \cite{BGK,Morse} and
the online enhancement of Fig. \ref{fig4} demonstrates their longevity and
their stability even when they cross the shock-like structures. The electron 
distribution in the intervals $400 < |x|/\lambda_D < 500$ just outside of 
this trapped electron population shows a spatial variation. This variation 
is caused by the free electrons that escape upstream. The current of the 
escaping electrons must be compensated by a return current of the incoming 
electrons, which gives rise to a change of the electron's mean speed along 
the x-direction. The incoming upstream electrons are accelerated towards the 
shock. 

A third interval $|x|/\lambda_D < 50$ in Fig. \ref{fig4}(b) coincides with 
the downstream region that is enclosed by both shock-like structures. The 
ion density in this interval exceeds $2n_0$ and additional electrons can be 
confined. The trapped electrons gain kinetic energy as they move into a 
region with a higher positive potential, which explains why their peak 
velocity is correlated to the ion density. The peak velocity is not reached 
by the electrons at $x\approx 0$ due to the negative potential of the ion 
phase space hole that is located at this position. The fastest electrons are 
found instead close to the shock-like structures at $|x|/\lambda_D \approx 
50$ where the potential peaks in Fig. \ref{fig3}. 

\begin{figure}
\includegraphics[width=8.2cm]{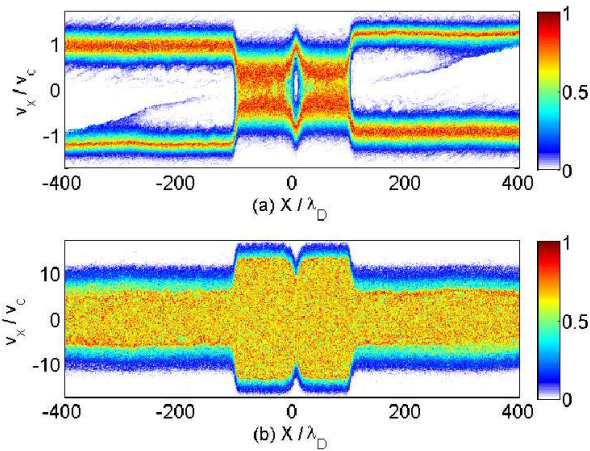}

\caption{The phase space distributions $f_{i,e}(x,v_x)$ from the 1D simulation
at the time $t\omega_{pi}=157$: Panel (a) shows the ion distribution. 
Panel (b) shows the electron distribution. Space and velocity are expressed 
in units of the Debye length $\lambda_D$ and of the initial cloud speed $v_c$. 
The density is normalized to its peak value and displayed on a linear color 
scale.}\label{fig5}
\end{figure}

Figure \ref{fig5} shows the phase space distributions of the ions and 
electrons at $t\omega_{pi} = 157$. The strong electrostatic fields in Fig. 
\ref{fig2} maintain the narrow transition layers in Fig. \ref{fig5}(a), 
which separate the downstream region with $|x|/\lambda_D <100$ from the 
foreshock regions of both shock-like structures. The ion beams at 
$x/\lambda_D > 100$ and $v_x \approx v_c$ and at $x/\lambda_D < -100$ and 
$v_x \approx -v_c$ in the displayed spatial interval consist now almost 
exclusively of ions that were reflected by the shock or accelerated upstream 
from the downstream region. The ion phase space density distribution in Fig. 
\ref{fig5}(a) does still not reach its peak value at $v_x = 0$ in the 
downstream region, which we would expect from a fully thermalized ion 
distribution. This aspect has been observed in previous simulations 
\cite{Parametric} that employed a different PIC simulation code and an 
ion-to-electron mass ratio of 400 rather than 250.

The electron distribution in Fig. \ref{fig5}(b) does again not show a 
Maxwellian velocity distribution in the displayed interval. The phase space
distribution shows a constant density at low speeds and a fast decrease for 
$|v_x / v_c|>7$ in both foreshock regions and for $|v_x / v_c|>17$ in the 
downstream region. The potential of the ion phase space hole, which is 
negative relative to that of the surrounding downstream region, 
continues to repel electrons, by which it decreases their peak speed at 
$x\approx 0$. The flat-top velocity distribution of the electrons converges 
to its initial Maxwellian distribution outside of the foreshock region. The 
similarity between the plasma distributions in Fig. \ref{fig4} and \ref{fig5} 
evidences that the shock-like structures are stationary in their rest frames 
in the considered case.

The 1D simulation demonstrates that the selected initial conditions result
in the growth and stable propagation of a pair of shock-like structures. 
However, the positive potential of the overlap layer is initially not 
sufficiently strong to reflect the incoming ions. The extra potential, 
which is needed for the shock formation, is provided by a gradual 
localized accumulation of ions during $t \omega_{pi} \approx 20$. 
This time delay has important consequences for the shock formation in 
more than one dimension, which is demonstrated by a direct comparison of 
the field distributions computed by the 1D and 2D simulations. 

\subsection{The 2D simulation}

\begin{figure}
\includegraphics[width=8.2cm]{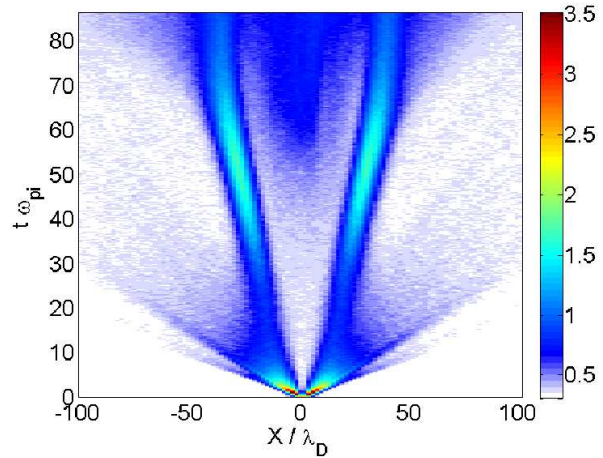}
\caption{The evolution of $10^3 {\langle E_{2D}^2 \rangle_y}^{1/2}$, where
$\langle E_{2D}^2 \rangle_y$ is the energy density of the in-plane electric 
field, which has been averaged along the y-direction. Space is normalized to 
the electron Debye length $\lambda_D$ and time is normalized to the ion 
plasma frequency $\omega_{pi}$. The color scale is linear.}\label{fig6}
\end{figure}

Figure \ref{fig6} visualizes the square root of the energy density $\langle 
E_{2D}^2 (x) \rangle_y = \frac{1}{300} \sum_{j=1}^{300} \left ( E_x^2 
(x,j\Delta_y) + E_y^2 (x,j\Delta_y) \right )$ of the in-plane electric field, 
which has been averaged along the y-direction. 
The field distribution evolves qualitatively similarly in the 2D simulation
and in the 1D simulation (See Fig. \ref{fig2}) until $t\omega_{pi} \approx 5$. 
The ion density is gradually increased beyond $2n_0$ in both simulations 
during $5 < t\omega_{pi} < 15$, but the ion compression instability has not 
yet resulted in strong electrostatic fields. 

The ion compression instability results in a visible field growth after 
$t\omega_{pi} \approx 25$ in both simulations. The unipolar electric fields, 
which sustain both shocks in the 1D simulation, saturate at around 
$t\omega_{pi} \approx 50$ in Fig. \ref{fig2} and maintain thereafter a 
constant peak amplitude. The energy density of the in-plane electric field 
in Fig. \ref{fig6} evolves qualitatively different after the time 
$t\omega_{pi} \approx 50$ when it reaches its maximum. The energy density 
of both pulses decreases and they slow down. The weakening of both pulses 
is accompanied by a rise of the field energy density in the interval they 
enclose. 

\begin{figure}
\includegraphics[width=8.2cm]{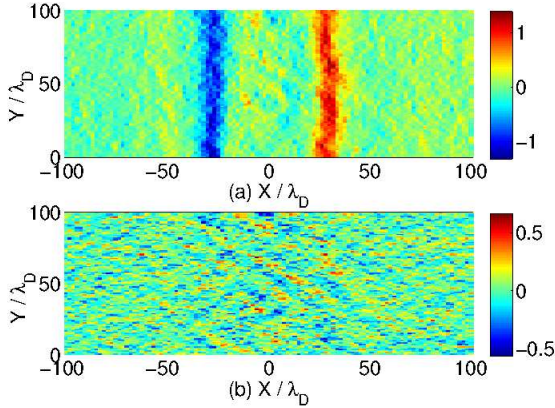}
\caption{The in-plane electric field at the time $t\omega_{pi}=50$:
The upper panel (a) shows $10^3 E_x(x,y)$ and the lower panel (b) shows 
$10^3 E_y (x,y)$.}\label{fig7}
\end{figure}

The in-plane components of the electric field at the time $t\omega_{pi}=50$ 
are shown in Fig. \ref{fig7}.
The $E_x$-component reveals unipolar electric field pulses 
at $|x|/\lambda_D \approx 30$ with a polarity that is typical for the 
ambipolar electric field. These field pulses put the interval $|x|/\lambda_D
<20$ on a positive potential relative to the surrounding plasma, which
helps confining the electrons. Weak coherent electric field patches are
visible within $|x|/\lambda_D <30$ in the otherwise noisy $E_y$-component. 
The field distribution is practically planar at this time and the plasma
dynamics should be analogeous to that in the 1D simulation. 

\begin{figure}
\includegraphics[width=8.2cm]{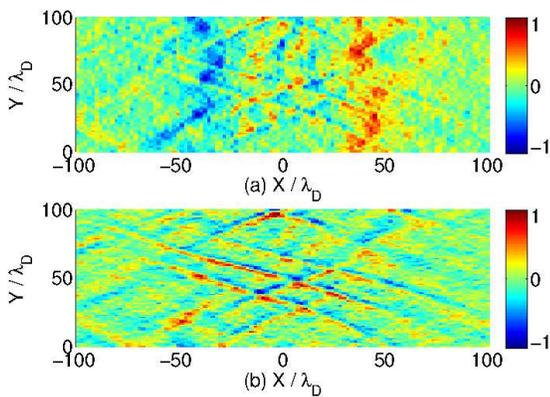}
\caption{The in-plane electric field at the time $t\omega_{pi}=86$:
The upper panel (a) shows $10^3 E_x(x,y)$ and the lower panel (b) shows 
$10^3 E_y (x,y)$.}\label{fig8}
\end{figure}

The electric field topology has changed significantly at the time 
$t\omega_{pi}=86$, which is evidenced by Fig. \ref{fig8}.
The amplitude of $E_x$ is only slightly lower than that in Fig. \ref{fig7}. 
The main difference compared to Fig. \ref{fig7}(a) is that the field 
distribution is no longer planar. Averaging the electric field energy density 
at $t\omega_{pi}=86$ like in Fig. \ref{fig6} results in a broader spatial 
interval with a lower energy density compared to that at $t\omega_{pi}=50$. 
The interval enclosed by both pulses shows oblique wave structures. The 
electric field is no longer planar and anti-parallel to the velocity vector 
of the incoming ions. The ions are thus not only slowed down along $x$, but 
they are also deflected along $y$ by the ambipolar electric field. This 
deflection changes the balance between the upstream pressure and the pressure 
of the plasma within the overlap layer, which is essential for a shock 
formation and stabilization.

\begin{figure}
\includegraphics[width=8.2cm]{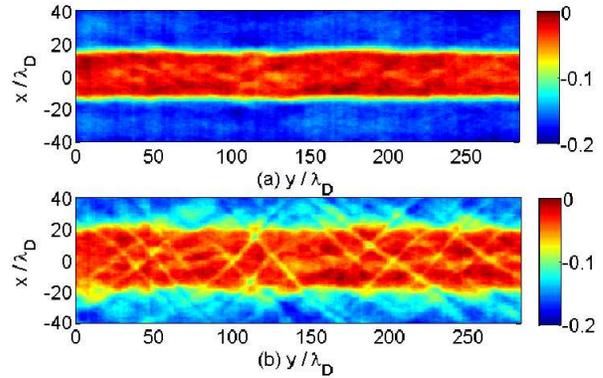}
\caption{The normalized electrostatic potential $\tilde{U}(x,y)$ computed 
by the 2D simulation at the time $t\omega_{pi}=50$ (a) and at $t\omega_{pi}
=86$ (b). The color scale is linear.}
\label{fig9}
\end{figure}

Figure \ref{fig9} depicts the electrostatic potentials close to $x=0$ 
of the field distributions at $t\omega_{pi}=50$ and $86$.
This potential $\tilde{U}(x,y)$ is computed in the same way and with the 
same normalization as the one shown in Fig. \ref{fig3}. The magnitude of
the potential difference between $x\approx 0$ and $|x|/\lambda_D \approx 
40$ is about 0.2 in both cases. The potential difference that sustains
the stable shock-like structures in the 1D simulation is 3-4 times larger 
and we expect clear differences between the plasma distributions in both 
simulations. The potential structure at $t\omega_{pi}=50$ is practically 
planar. It is more diffuse at $t\omega_{pi} =86$ and we observe oblique 
structures within the high potential region.

We examine the projection of the phase space density distributions of 
electrons and ions onto the $(x,v_x)$ plane in form of an animation and at 
selected time steps. The phase space distributions of electrons and ions are 
integrated over the y-direction. The purpose of examining the phase space 
density distributions is to better understand the time-evolution of the ion 
compression instability and the conditions, under which the ion acoustic 
instability can grow. The integrated phase space density distributions will 
also reveal differences caused by the dissimilar electrostatic potentials in 
the 1D and 2D simulations. We discuss the plasma phase space distribution at 
$t\omega_{pi}=10$ when the overlap layer has developed while $E_x$ is still 
weak in Fig. \ref{fig6}, at $t\omega_{pi}=50$ when the electric fields 
driven by the ion compression instability reach their peak amplitude and at 
$t\omega_{pi}=86$.

\begin{figure}
\includegraphics[width=8.2cm]{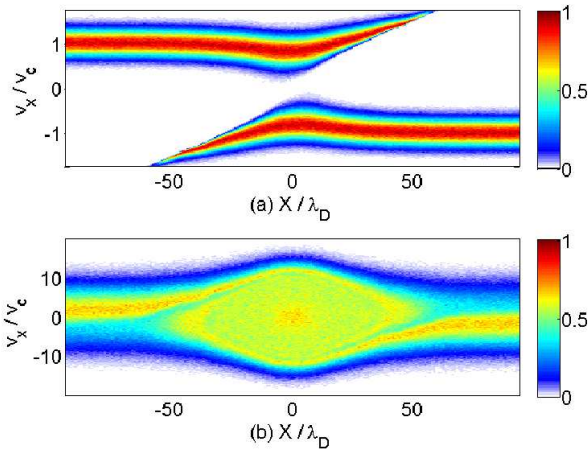}
\caption{The y-integrated plasma phase space distributions $f_{i,e}(x,v_x)$ 
at the time $t\omega_{pi}=10$: Panel (a) shows the ion distribution and panel
(b) the electron distribution. Space and velocity are normalized to the
electron Debye length $\lambda_D$ and the cloud speed $v_c$. The density
is normalized to the peak value reached in the simulation and the color 
scale is linear (enhanced online).}
\label{fig10}
\end{figure}

The ion phase space distribution in Fig. \ref{fig10}(a) shows some
modifications, which were not captured by our simple model of the overlap 
layer depicted in Fig. \ref{fig1}. The ions of both clouds have 
interpenetrated in the interval $-55 < x/\lambda_D < 55$. Their mean velocity 
modulus has decreased below $v_c$ at $x\approx 0$, where it has its minimum. 
Consider the ion beam located in the left half of the simulation box, which 
moves at a positive speed to the right. As these ions approach the overlap 
layer, they experience its repelling electrostatic potential. They are 
accelerated again by the electric field in the interval $x>0$. Some of the 
ions at the front $x / \lambda_D \approx 55$ and $v\approx 1.7 v_c$ have 
reached a speed that is higher than that of any ion in the initial 
distribution. These ions entered the overlap layer before the ambipolar 
electric field could build up and, hence, they were not slowed down by it. 
By the time they leave the overlap layer the electric field has developed and 
the ions are accelerated. This acceleration is strongest at early times (See 
online enhancement of Fig. \ref{fig10}, which 
animates the phase space evolution for $0\le t \omega_{pe} \le 86$), when 
the ion density gradient and, thus, the ambipolar electric field are large. 
They have gained kinetic energy at the expense of electron energy in the 
time-dependent potential of the overlap layer.  

The ion beam fronts are no longer parallel to the $v_x$ direction. The faster 
the ions the farther they have propagated during the time interval $t\omega_p 
= 10$. The shear of the ion beam front is thus caused by the velocity spread 
of the ions, which corresponds to diffusion. This diffusion decreases the 
magnitude of the ion density gradient between the overlap layer and the 
incoming plasma and thus the amplitude of the ambipolar electric field. 
Diffusion is responsible for the observed rapid decrease of the electric 
field amplitude at early times in Fig. \ref{fig6}. 

The electron distribution in the online enhancement of Fig. \ref{fig10}(b) 
shows initially a spiral close to $x=0$ that is brought about by electron 
trapping in the growing potential of the expanding overlap layer. The 
electrons would form a vortex in a stationary positive potential. The spiral 
forms because firstly the entry points of the electrons into the overlap 
layer move in time to larger values of $|x|$ and, secondly, because the 
potential difference between the overlap layer and the surrounding plasma 
increases in time. Electrons that enter the overlap layer at a later time 
thus get accelerated to a larger speed. The increase of the potential is, in 
turn, a consequence of the ion compression due to their decreasing mean speed 
in Fig. \ref{fig10}(a).

Like in the 1D simulation, the current due to the electrons that leave the 
overlap layer drives an electric field just outside of the overlap layer. 
The electrons at $x/\lambda_D \approx -50$ are accelerated to positive $v_x$ 
by this electric field and they are thus dragged towards the overlap layer. 
More electrons flow towards the overlap layer than away from it. The net flux 
of electrons into the overlap layer is a consequence of its expansion in 
time, which implies that its overall ion number increases. The fastest 
electrons do not follow the shape of the trapped electron structure. Electrons 
entering at $x/\lambda_D = -95$ with $v_x = 10v_c$ in Fig. \ref{fig10}(b) are 
accelerated by the positive potential of the overlap layer as they approach 
$x=0$ and they are decelerated again as they move to larger positive $x$. 
These electrons are free.

\begin{figure}
\includegraphics[width=8.2cm]{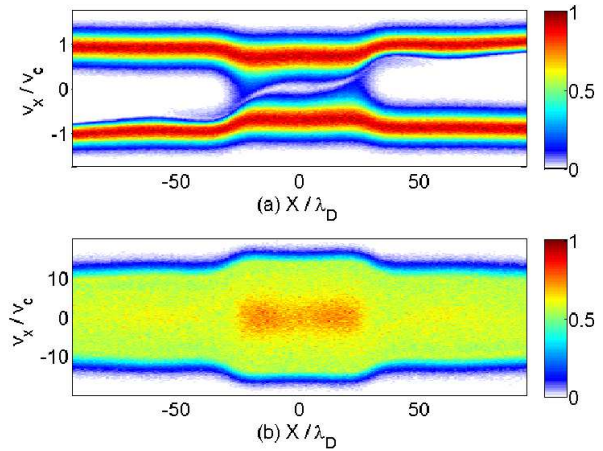}
\caption{The y-integrated plasma phase space distributions $f_{i,e}(x,v_x)$ 
at the time $t\omega_{pi}=50$: Panel (a) shows the ion distribution and panel
(b) the electron distribution. Space and velocity are normalized to the
electron Debye length $\lambda_D$ and the cloud speed $v_c$. The density
is normalized to the peak value reached in the simulation and the color 
scale is linear.}
\label{fig11}
\end{figure}

Figure \ref{fig11} shows the plasma phase space distribution at the time
$t\omega_{pi} = 50$. The overlap layer has expanded from $|x|/\lambda_D = 
55$ to the position $|x|/\lambda_D \approx 300$, which is outside of the 
displayed interval. A direct comparison of the Figs. \ref{fig10}(a) and 
\ref{fig11}(a) shows one difference between the ion distributions. Both 
ion beams are slowed down at the same position $x \approx 0$ at $t\omega_{pi} 
= 10$. They are decelerated most at $x/\lambda_D \approx \pm 30$ at 
$t\omega_{pi}=50$. Both points of maximum ion slow-down and compression are 
separated in space and enclose a region of enhanced ion density. The 
electrostatic fields, which are responsible for the ion slow-down at 
$|x|/\lambda_D \approx 30$, are sufficiently strong to reflect a fraction 
of the incoming ions at these locations. This can be seen more clearly in 
the animation (online enhancement of Fig. \ref{fig10}).

The phase space distribution of the electrons in Fig. \ref{fig11}(b) is
determined by the electrostatic potential set by the ion density, which is 
compressed beyond the value $2n_0$ in the interval $-30 < x/\lambda_D < 30$. 
One feature of the electron distribution that sets it apart from its
counterpart in the 1D simulation (See Figs. \ref{fig4}(b) and \ref{fig5}(b)) 
is that it is not a flat top distribution at low speeds. A weak enhancement 
of the phase space density can be observed at $v_x \approx 0$ in the interval 
$-25 < x / \lambda_D < 25$. The online enhancement of Fig. \ref{fig10} shows 
that the electron's phase space density in this interval continues to grow 
after $t\omega_{pi}\approx 50$. It is thus temporally correlated with the 
rise of the energy density close to $x\approx 0$ in Fig. \ref{fig6}. 

\begin{figure}
\includegraphics[width=8.2cm]{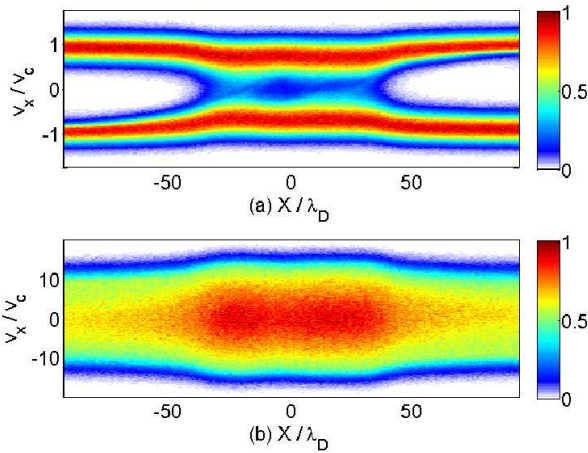}
\caption{The y-integrated plasma phase space distributions $f_{i,e}(x,v_x)$ 
at the time $t\omega_{pi}=86$: Panel (a) shows the ion distribution and panel
(b) the electron distribution. Space and velocity are normalized to the
electron Debye length $\lambda_D$ and the cloud speed $v_c$. The density
is normalized to the peak value reached in the simulation and the color 
scale is linear.}\label{fig12}
\end{figure}

Figure \ref{fig12} depicts the plasma phase space distributions at 
$t\omega_{pi} = 86$. 
The large scale distribution of the ions in the 2D simulation resembles
that in the 1D simulation in Fig. \ref{fig4}(a) (not shown) except in the 
interval displayed in Fig. \ref{fig12}(a). We observe an overlap layer with
two dense counter-streaming ion beams and a dilute ion population with
$|v_x| \approx 0$. The online enhancement of Fig. \ref{fig10} shows 
that the velocity gap between both dense ion beams increases again after 
$t\omega_{pi} \approx 50$, while both ion beams converged along the 
$v_x$-direction in the 1D simulation. The plasma has thus evolved to a 
different nonlinear state at this time in the 1D and 2D simulations. The 
counter-streaming ion beams in the 2D simulation are affected significantly 
less by the positive potential of the overlap layer than those in the 1D 
simulation, which is a consequence of the different magnitude of the 
potential. Most ions in the 2D simulation experience the overlap layer as 
a localized potential maximum, which is not strong enough to slow them 
down to the ion's thermal speed in the downstream reference frame. The
velocity change of the bulk ions close to $|x|/\lambda_D \approx 50$ is
of the order of $v_c/3$, which is comparable to or below the sound speed 
$c_s$. 

An ion distribution, which is symmetric around $v_x =0$, corresponds to a 
hybrid structure with equally strong electrostatic shock and double layer 
components. The ion distribution in Fig. \ref{fig4}(a) is less symmetric
than that in Fig. \ref{fig12}(a). The ion beam in the interval $50 < x /
\lambda_D < 100$ and $v_x > 0$ in Fig. \ref{fig4}(a), which is composed 
of trapped incoming ions and of ions that are accelerated from the downstream 
region into the upstream direction, is significantly thinner than the incoming 
free ion population with $v_x < 0$. The hybrid distribution in the 1D 
simulation thus has a much stronger electrostatic shock character than its 
counterpart in Fig. \ref{fig12}(a) at this time.

The phase space distribution of the electrons in Fig. \ref{fig12}(b) shows a 
pronounced maximum at $v_x \approx 0$ in the interval $-50\le x/\lambda_D \le 
50$. It is closer to a Maxwellian than to a flat-top velocity distribution. 
We attribute the differences between the electron distributions in Fig. 
\ref{fig4}(b) and \ref{fig12}(b) to the higher-dimensional phase space 
dynamics in the 2D simulation. The electron dynamics is confined to the 
$(x,v_x)$ plane in the 1D simulation. The oblique electric fields observed 
in Fig. \ref{fig8} introduce an electric force component in the y-direction 
that is a function of both spatial coordinates. The phase space dynamics of
the electrons involves in this case the four coordinates $(x,y,v_x,v_y)$. The 
growing amplitudes of the ion acoustic waves (Compare Figs. \ref{fig7} and
\ref{fig8}) imply that they can interact nonlinearly with electrons in a 
velocity interval that increases in time. 

This discrepancy between the electron phase space distributions in the 1D 
and 2D simulations reveals another reason for why the Mach number is not as 
meaningful in a kinetic collision-less framework as it is in a collisional 
fluid theory. The adiabatic index $\gamma_s$ is tied to the degrees of 
freedom in the medium under consideration. The particles of the mono-ionic 
plasma in the PIC simulation have three degrees of freedom. However, only 
one degree of freedom is accessible to particles in a 1D simulation of 
electrostatic processes or in the 2D simulation, if the electrostatic fields 
are perfectly planar. The onset of the ion acoustic instability makes 
accessible a second degree of freedom to the plasma and $\gamma_s$ can change.  

\begin{figure}
\includegraphics[width=8.2cm]{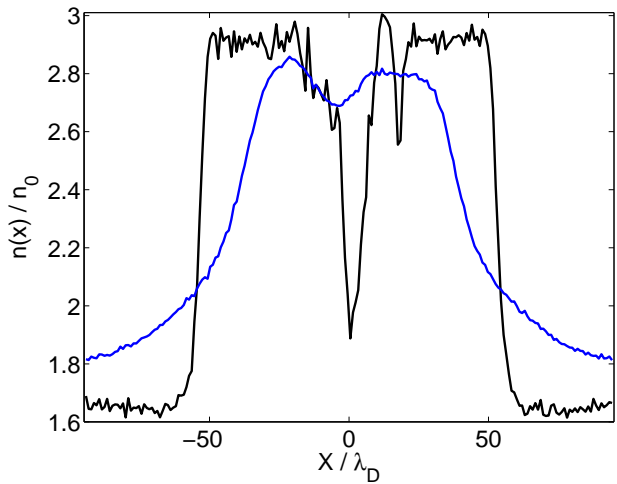}
\caption{The y-integrated ion density distributions in the 1D simulation 
(black curve) and in the 2D simulation (blue curve) at $t\omega_{pi}=86$.}
\label{fig13}
\end{figure}

The ion density distributions $n(x) = \int_{-\infty}^\infty f_i (x,v_x) dv_x$ 
computed from the y-integrated phase space distributions Fig. \ref{fig4}(a) 
and Fig. \ref{fig12}(a) shed further light on the different plasma state in
the 1D and 2D simulations. Figure \ref{fig13} compares both distributions at 
$t\omega_{pi}=86$.
The ion density distribution in the 1D simulation shows steep gradients
between the downstream region and the foreshock regions of both shock-like
structures. The ion density grows from the foreshock value $n(x)\approx 1.65 
n_0$ close to $|x|/\lambda_D \approx 60$ to the downstream value $n(x)\approx 
2.9$ over $10 \lambda_D$. The ion cavity at $x\approx 0$ is caused by the ion 
phase space hole. The ion density gradient in the 2D simulation is lower and 
the peak density is reached at $|x|/\lambda_D \approx 20$, which is well 
behind the shock location in the 1D simulation. The wide transition layer in 
the 2D simulation is partially a consequence of averaging the ion density over
the y-direction; the potential distribution in Fig. \ref{fig9} demonstrates
that the overlap layer is not perfectly planar at this time. Another important
reason for the wide transition layer is that the ion beams in Fig. 
\ref{fig12}(a) are slowed down less and over a wider spatial interval than
the ion beams in Fig. \ref{fig4}(a), which results according to the continuity
equation in a lower density gradient.

We have observed significant differences in the plasma evolution in the
1D and 2D simulations during the time interval $50 \le t\omega_{pi} \le
86$ (Compare Figs. \ref{fig2} and \ref{fig6}). We have attributed these
difference to the oblique electrostatic structures in Fig. \ref{fig9} that 
are geometrically suppressed in the 1D simulation. Their obliquity suggests 
that they are driven by an ion acoustic wave instability between the two 
ion beams, which counter-stream at a speed that exceeds the ion acoustic 
speed \cite{ForslundC}. Their growth time is of the order of ten inverse
ion plasma frequencies, which suggests that the instability is ionic. 

\begin{figure}
\includegraphics[width=8.2cm]{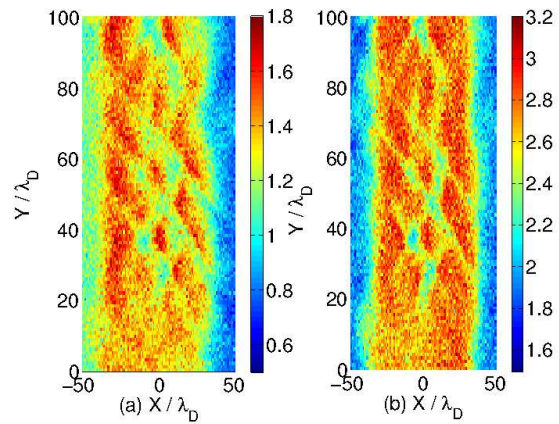}
\caption{The ion density distributions in a section of the 2D simulation 
box at the time $t\omega_{pi}=72$. Panel (a) shows the distribution
of the ion beam that moves to increasing values of $x$. Panel (b) shows
the total ion density (enhanced online).}\label{fig14}
\end{figure}

We turn towards the ion density distribution in the 2D simulation as a means 
to determine whether or not the ion acoustic instability is involved and if 
it is indeed responsible for the different ion evolution in both simulations. 
The ion acoustic instability is purely growing (the wave frequency has no 
real part) for our symmetric beam configuration \cite{ForslundC}. Its phase 
speed vanishes. We thus expect the growth of spatially stationary oblique 
ion density modulations in the overlap layer. The presence of such structures 
is confirmed by Fig. \ref{fig14}, which shows the ion density distribution at 
$t\omega_{pi}=72$ 
(The online enhancement of Fig. \ref{fig14} animates the ion density evolution 
until $t\omega_{pi}=72$). The ion distribution is initially planar. The 
online enhancement shows the formation of the overlap layer (See Fig. 
\ref{fig10}), which is followed by a compression phase that results  
in a planar ion pile-up. The density of the left ion beam (panel (a) in the 
online enhancement) increases initially at $x/\lambda_D \approx -30$ (See 
also Fig. \ref{fig11}). Eventually a filamentation of the single beam can be 
observed while the total ion density remains spatially uniform. The ion 
acoustic instability thus separates the ion beams in the direction that is 
orthogonal to their flow direction but it leaves the total density unchanged. 
The filaments do not move in the x-y plane as they develop, which implies
that the waves tied to them have a vanishing phase speed. The total ion 
density is modulated at late times as well (See Fig. \ref{fig14}(b)), which 
results in the electrostatic fields that are strong enough to modulate the 
potential of the overlap layer in Fig. \ref{fig9}. 

The ion acoustic waves yield spatial modulations of the ion density, which
are of the order of $n_0/10$ and they result in oblique ion flow channels 
in Fig. \ref{fig14}(a). Their electric fields are thus strong enough to 
deflect the ions in the $x,y$-plane, which is at least partially responsible 
for the diffuse ion population with $v_x \approx 0$ in Fig. \ref{fig12}(a). 
The number density of this diffuse ion population is significantly less than 
the density $n_0$ of each beam. However, we have to compare the number 
density of the diffuse ion component with the change of the ion number 
density, which is imposed by the beam velocity change. The latter is 
significantly less than $n_0$. This explains why the peak density of the ions 
in Fig. \ref{fig13} is comparable in both simulations even though the phase 
space distributions in Fig. \ref{fig4}(a) and \ref{fig12}(a) differ 
significantly. The online enhancement of Fig. \ref{fig10} also shows that the
velocity change of the ion beams is reduced as the diffuse ion beam component 
forms. We infer that the ion acoustic instability is indeed responsible for
the change of the character of the beam overlap layer in the 1D and 2D 
simulations.

\section{Discussion}

We have examined here the interplay of the ion compression instability, 
which triggers the formation of a non-relativistic electrostatic shock, and 
the ion acoustic instability. The ion acoustic waves can not grow if the 
speed modulus of the ion beams exceeds the ion acoustic speed. Ion acoustic 
waves can thus only grow for the initial conditions considered here, 
if their wave vector is oblique to the flow direction. The projection of the 
ion velocity onto the wave vector is in this case subsonic and the ion beams 
can couple to the waves \cite{ForslundC}. The ion acoustic instability is 
alike its relativistic counterpart \cite{Gedalin}, which results in the 
aperiodic growth of strong magnetowaves. The low flow speeds, which we 
examine here, imply that electrostatic forces remain stronger than the 
magnetic ones and the waves are electrostatic. The ion compression instability 
and the ion acoustic instability can thus be distinguished by the orientation 
of the wave vector of their electric field relative to the flow direction.

Their simultaneous growth is made possible by a delayed formation of the
shock-like structures. We have defined a 
shock-like structure as a combination of electrostatic shocks and double 
layers as discussed in Ref. \cite{GreatPaper}. Shock-like structures 
evolve into electrostatic shocks once the downstream region is sufficiently
large to thermalize the ion distribution, which reduces the number of ions
that can reach the shock and be accelerated into a double layer structure. 
The time that it takes to form a pair of such 
structures out of the collision of two identical plasma clouds is influenced 
by how the cloud collision speed compares to the ion acoustic speed $c_s$. 
They form on electron time scales if the Mach number of the 
cloud collision speed $2v_c$ is about 2-3 and on ion time scales if it is 
$\approx 4$ \cite{Parametric}. This difference arises because the upstream 
ions can be slowed down directly to downstream speeds by the ambipolar 
electric field between the plasma overlap layer and the upstream plasma in 
the first case. In the second case, the ion compression instability has to 
pile up the ions to increase the potential difference between the overlap 
layer and the upstream plasma to the value that is required for the creation 
of shocks. The ion compression instability becomes inefficient for much 
larger collision Mach numbers than 4 \cite{MildRel}, at least for the initial 
conditions we have selected here. 

Shocks driven by rarefaction waves \cite{SarriPRL} may have other limitations. 
A collision of clouds with unequal densities can increase the maximum Mach 
number up to which shocks can form \cite{Sorasio}. Faster shocks can also 
form after beam instabilities have developed, which either increase the 
amplitude of the ambipolar electric field through electron heating 
\cite{Sorasio} or provide additional stabilization by self-generated 
magnetic fields \cite{Antoine,Pohl,Kazimura,Spitkovsky}.    

Our results are as follows. A 1D simulation, which employed the 
ion-to-electron mass ratio 250 and the fastest Mach number that resulted in 
the formation of shock-like structures, confirmed that this formation is 
delayed by tens of inverse ion plasma frequencies.
The time it takes the shock-like structures to form is comparable to that 
obtained for a mass ratio 400 \cite{Parametric}. This delay thus does not 
seem to be strongly dependent on the ion mass, as long as it is sufficiently 
high to separate electron and ion time scales. 

This time delay has important consequences in a 2D simulation, which permits 
the ion acoustic instability to develop. The short-wavelength structures 
generated by the ion acoustic instability in the overlap layer, which is the 
region where the ions of both plasma clouds interpenetrate, break the 
planarity of the electrostatic wave fronts driven by the ion compression 
instability and the wave fields become patchy. A fraction of the ions is 
thermalized as they enter the overlap layer with its strong ion acoustic 
waves and they form a diffuse ion component with a low velocity along the 
cloud collision direction. This diffuse ion population thus expands only 
slowly. Its density modifies the character of the shock-like structures. 
These structures were closer to electrostatic shocks in the 1D simulation, in 
which no diffuse ion component formed, while the double layer component and 
the electrostatic shock component were almost equally strong in the 2D 
simulation with the diffuse component. 

The ion density reached a similar peak value in both simulations but the 
transition layer of the shock-like structures in the 2D simulation has been 
significantly broader than that in the 1D simulation. The ion acoustic 
instability does thus not only affect the stability and the structure of 
the transition layer of an existing electrostatic shock 
\cite{Karimabadi,Kato}, but also its formation.

Our results indicate so far that the ion acoustic instability reduces the 
maximum Mach numbers that can be reached by stable electrostatic shocks 
with a narrow Debye length-scale transition layer to values below the 
limit obtained from one-dimensional models or simulations. The shock-like 
structures form faster at lower Mach numbers of the collision speed and 
the ion compression instability can outrun the ion acoustic instability;
the shock should in this case be similar to the one in our 1D simulation. 
A difference in the structure of the shock transition layer may have 
consequences for experiments, which detect electric field distributions 
in plasma. An example is the proton radiography method \cite{SarriPRT}. 
Shocks with a narrow transition layer result in much stronger and spatially 
confined electric fields. The shock we observe in the 2D simulation yields 
diffuse and weaker electric fields. Such field distributions may in some 
cases not be associated with electrostatic shocks.

{\bf Acknowledgements:} M.E.D. wants to thank Vetenskapsrådet for financial 
support. M.P. acknowledges support through Grant PO 1508/1-1 of the Deutsche 
Forschungsgemeinschaft (DFG). The computer time and support has been provided 
by the High Performance Computer Centre North (HPC2N) in Umeå.

\end{document}